\documentclass[twocolumn,preprintnumbers,prl,showpacs,amsmath,amssymb]{revtex4-2}

\usepackage{graphicx}
\usepackage{dcolumn}
\usepackage{bm}

\usepackage{textcomp}
\usepackage{bm}
\usepackage{setspace}
 \usepackage[normalem]{ulem}
 \useunder{\uline}{\ul}{}

\AtBeginDocument{%
  \addtolength\abovedisplayskip{-0.7\baselineskip}%
  \addtolength\belowdisplayskip{-0.7\baselineskip}%
}
\newcommand{\ber}{\begin{eqnarray}}
\newcommand{\eer}{\end{eqnarray}}
\newcommand{\bea}{\begin{equation}}
\newcommand{\eea}{\end{equation}}

\begin{document}

\title{On temperature discontinuity at an evaporating water interface}

\author{Parham Jafari}
\author{Amit Amritkar}
\author{Hadi Ghasemi}
\email{Electronic address: hghasemi@uh.edu}
\affiliation{University of Houston\ Department of Mechanical Engineering,\ 4726 Calhoun Road,\ Houston, TX, USA, 77204-4006}
\date{\today}
             

\begin{abstract}
Evaporative mass flux is governed by interfacial state of liquid and vapor phases. For closely similar pressures and mass fluxes of liquid water into its own vapor, discontinuity between interfacial liquid and vapor temperatures in the range of 0.14-28 K is reported. This controversial discontinuity has resulted in an obstacle on understanding and theoretical modeling of evaporation. Here, through study of vapor transport by Boltzmann transport equation solved through Direct Simulation Monte Carlo Method, we demonstrated that the measured discontinuities were strongly affected by boundary condition on the vapor side of the interface and do not reflect the interfacial state. The temperature discontinuity across the evaporating interface is $\le$ 0.1 K for all these studies. To accurately capture the interfacial state, the vapor heat flux should be suppressed. 

\end{abstract}

\maketitle
Evaporation phenomenon is the governing pillar of a wide range of disciplines ranging from atmospheric sciences to energy and biology. Kinetic of evaporation is described by various theories including diffusion \cite{Maxwell1890,HisatakeTanakaAizawa,Hu2002,SefianeBennacer}, Hertz-Knudsen (HK) \cite{Hertz,Knudsen1,Bond2004}, Statistical Rate Theory (SRT)\cite{WardFindlayRizk,Ward1999,FangWard2} and Non-Equilibrium Thermodynamics (NET)\cite{Bedeaux1999,Kjelstrup2008}. In all these theories, the kinetic of evaporation is governed by the interfacial thermodynamic properties (i.e. temperature and pressure) which are difficult to measure. Fang and Ward \cite{FangWard} conducted an accurate measurement of interfacial temperature of liquid and vapor at an evaporating water interface and found that a temperature discontinuity exists across the interface with the magnitude of up to 7.8 K. This was in contrast to all the previous measurements that considered approximately local equilibrium condition at the water interface \cite{ShankarDeshpande}. This contrast in temperature discontinuity brought an unprecedented hurdle on fundamental understanding of evaporation. Possible factors affecting measurement of the interfacial temperature discontinuity, including radiation and evaporative cooling of the thermocouple bead, were closely examined and concluded to be negligible. Various scientists conducted these experiments and reported temperature discontinuity of 0.14-28 K \cite{Jafari2018,Badam,BadamKumarDurstDanov,Persad2016,Polikarpov2019,Kazemi2017,Pao1971a,Persad2016,Ghasemi2011,Duanthesis,DuanBadamDurstWard,Ya2017,You2001,Ghasemi2010,PopovMellingDurstWard,Bond2004,McGaugheyWard,Zhu2013a,Zhu2009,FangWard,WardStanga,DuanThompsonWard,Kon2016,HolystLitniewski,BuffoneSefiane,Kazemi2017a}. Although majority of experiments indicated that liquid side of the interface is colder than the vapor side, few experiments \cite{Zhu2009,Zhu2013a} showed the opposite direction of temperature discontinuity. This temperature discontinuity at an evaporating water interface remains still a mystery. 

Here, we propose a molecular insight on the evaporation phenomenon and elucidate source of the mystery. This insight explains all the contradicting measurements conducted by various groups and provides a platform for further advancement of evaporation theories. The interface is only a few molecular length thick and determination of thermodynamic state on each side of the interface is difficult. One way to avoid the experimental challenges is to computationally analyze vapor transport in the Knudsen layer (Kn) and vapor phase. This layer forms during evaporation between a liquid surface and the bulk vapor phase, Fig. \ref{Fig1}. Thickness of this layer is in the order of a few molecular mean free path (mfp) which is written as $\lambda = \frac{k_b T}{\sqrt{2} P\pi d^2}$, where $k_b$ is the Boltzmann constant, $T$ is the vapor temperature, $P$ is the vapor pressure and $d$ is the molecular diameter of vapor phase, if they are approximated as hard spheres \cite{DuanWardBadamDurst}. When mfp is on the same order as the characteristic transport length scale (0.1$\le$Kn$\le$10), transport of vapor molecules in the Knudsen layer reveals a mixture of diffusive and ballistic characteristics referred as transitional transport which could be understood through solution of Boltzmann transport equation (BTE). We adopted Direct Simulation Monte Carlo (DSMC) method to solve BTE at an evaporating water interface. This method allows to accurately capture thermal field in the vapor side of the evaporating interface and determine temperature discontinuity.  
\vspace{-0.8 cm}
\section{Statement of the problem}
\vspace{-0.4 cm}

\begin{figure}[t]
\begin{center}
\includegraphics[width=0.38 \textwidth]{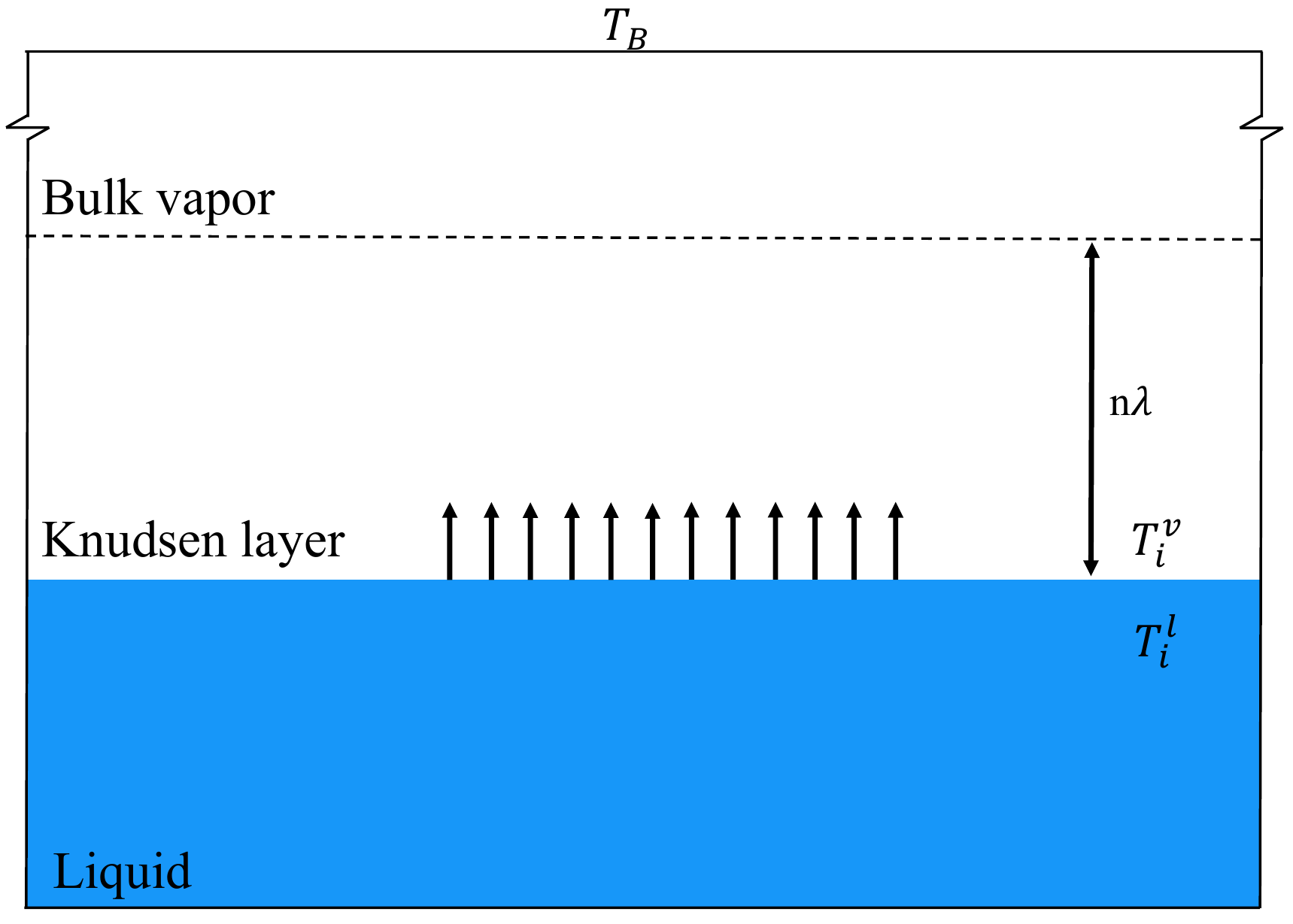}
\caption{Schematic of a planar evaporating interface, zoomed into the scale of the Knudsen layer. There are indeed three regions consist of liquid phase, Knudsen layer and the bulk vapor phase. The thickness of Knudsen layer is equal to a few molecular mean free path. $T_i^l$ and $T_i^v$ are temperatures at the liquid and vapor sides of the interface, respectively.}
\label{Fig1}
\end{center}
\end{figure}

Figure \ref{Fig1} shows an evaporating liquid-vapor interface and the interfacial temperatures which are denoted by $T_i^l$ and $T_i^v$ at the liquid and vapor sides of the interface, respectively. In the Knudsen layer, the vapor molecules are highly influenced by the evaporating interface and are in a non-equilibrium state. Above the Knudsen layer, there exists the bulk vapor phase with a boundary at temperature of $T_B$. The coordinate of this boundary is chosen as furthest reported temperature from liquid-vapor interface in the experimental literature. However, the coordinate of this boundary is arbitrary and does not affect the conclusions. 

\begin{figure}[!ht]
 \begin{center}
\includegraphics[width=0.49 \textwidth]{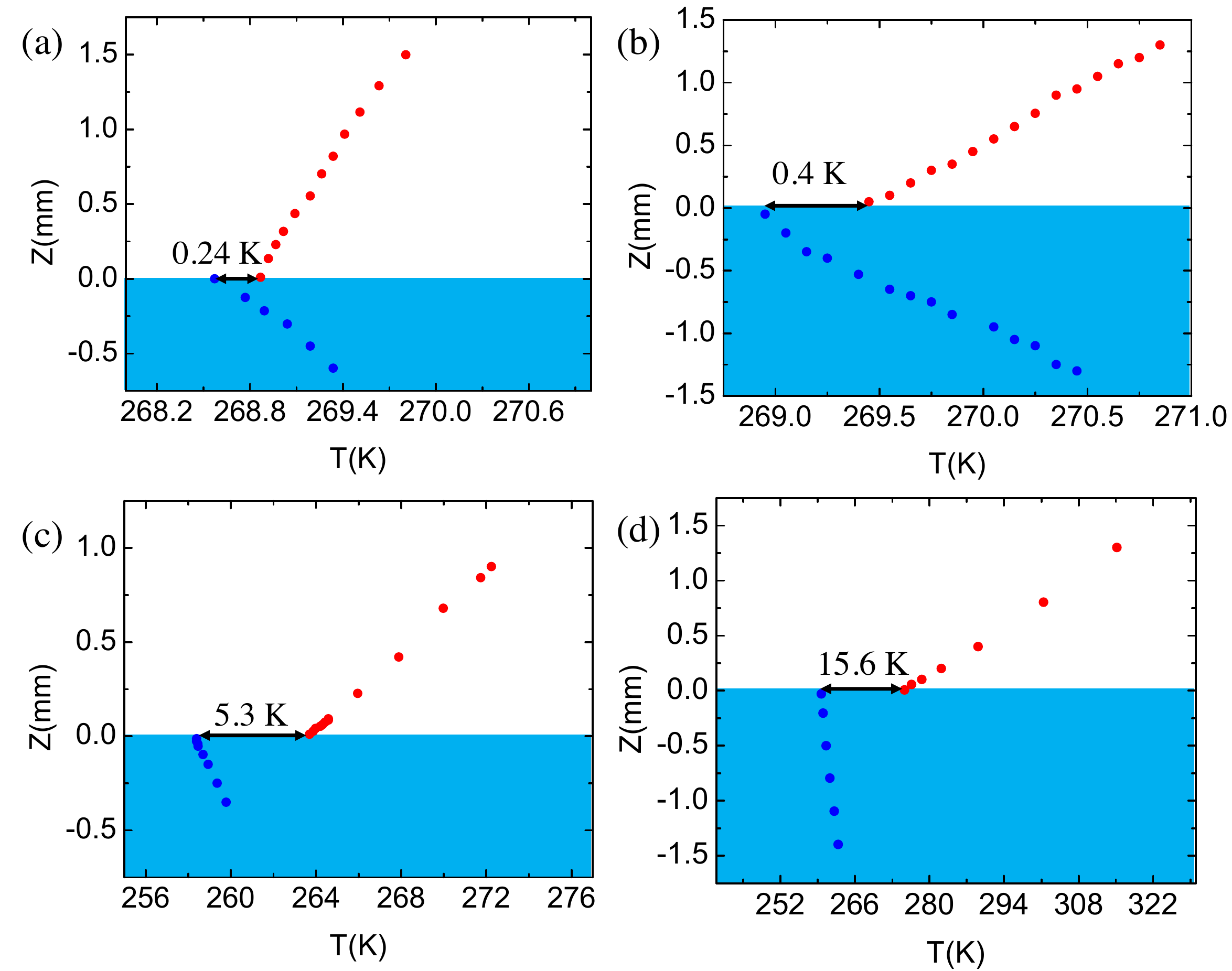}
\caption{Temperature profile in the liquid and vapor phases of an evaporating water into its own vapor from four independent groups. All these temperature profiles are measured at the centerline of the liquid-vapor system. Due to symmetry at the centerline, the role of convection is negligible. (a) Kazemi et al. \cite{Kazemi2017}. (b) Jafari et al. \cite{Jafari2018}. (c) Duan et al. \cite{DuanWardBadamDurst}, and (d) Badam et al. \cite{BadamKumarDurstDanov}. In (a), (b) and (c) studies, there was no direct heating element in the vapor phase, while in study (d), the authors used a mounted heating element with temperature of 80 \textcelsius\ above the free liquid surface.}
\label{Fig2}
\end{center}
\end{figure}

Figure \ref{Fig2} illustrates liquid and vapor side temperature profiles for water evaporation into its own vapor from four independent groups. Despite close evaporation rates, the measured temperature discontinuities by these groups varies in a wide range (i.e. 0.24  to 15.6 K). Kazemi et al. \cite{Kazemi2017} measured temperature discontinuity of 0.24 K at vapor pressure of 435 Pa and mass flux of 2.41$\times$10$^{-4}$ kg/(m$^2$s). In the experiments, there was no heating element in the vapor phase. Jafari et al. \cite{Jafari2018} measured temperature discontinuity of 0.4 K at 446 Pa and evaporative mass flux of 3.1$\times$10$^{-4}$ kg/(m$^2$s). In this experiment, the liquid container was mounted on a heating stage with temperature of 40 \textcelsius\ to increase heat flux to the interface. Similar to Kazemi et al. \cite{Kazemi2017}, there was no heating element in the vapor side of the interface. In a work by Duan et al.\cite{DuanWardBadamDurst} on water evaporation with vapor pressure of 176 Pa and the mass flux was 8.65$\times$10$^{-4}$ kg/(m$^2$s), interfacial temperature discontinuity of 5.3 K was reported. In these studies, the liquid bottom was kept at 277 K to suppress buoyancy convection.  Badam et al. \cite{BadamKumarDurstDanov} reported even greater interfacial temperature discontinuity of 15.6 \textcelsius\ for the case of 213 Pa vapor pressure and 12.3$\times$10$^{-4}$ Kg/(m$^2$s) evaporative mass flux. In this experiment, a heating element was mounted on the vapor side at coordinate of 3 mm above the liquid-vapor interface to boost heat flux to the liquid-vapor interface. 

\begin{figure}[!ht]
 \begin{center}
\includegraphics[width=0.43 \textwidth]{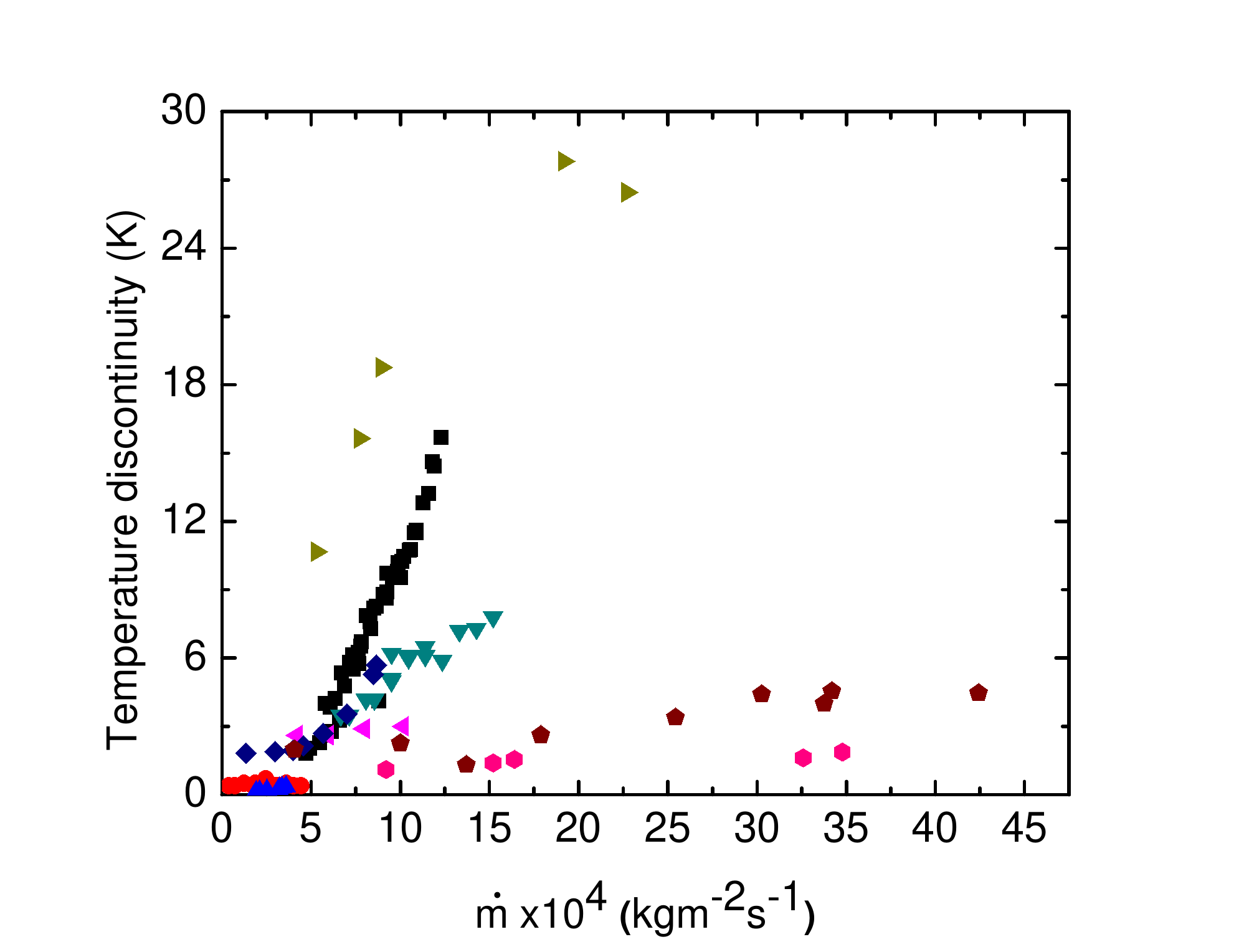}
\caption{Temperature discontinuity in various studies for a wide range of evaporation mass fluxes is shown. Note that for a given mass flux, the reported interfacial temperature discontinuities varies by two orders of magnitude.}
\label{Fig3}
\end{center}
\end{figure}

In Fig. \ref{Fig3}, all the reported temperature discontinuities for evaporating water are plotted as a function of mass flux. The reported values varies from 0.14 $\pm$ 0.1 K \cite{Kazemi2017} to more than 28 K $\pm$ 0.06 K \cite{Badam}. Based on the evaporation theories, mass flux is function of interfacial temperatures and pressures. Even for a similar set of pressure and mass flux, multiple values of temperature discontinuity are reported. As all these experiments are conducted with high accuracy, there should be a reason for these orders of magnitude variation in the reported temperature discontinuities. The main difference among these studies is the thermal boundary condition at the vapor side of the interface, $T_B$. Larger values of temperature discontinuity are associated with studies in which the vapor phase is heated with a heating element and smaller values of temperature discontinuity are reported in the case of no vapor heating.  Reported interfacial state of vapor is employed along with above equation to determine mfp values for different experimental conditions studied in literatures. Mfp varies from 5-28 $\mu$m in different studies and these mfps are mostly smaller or in few cases equal to the size of measurement probe (i.e. thermocouples). 
\vspace{-0.7cm}
\vspace{-0.9 cm}
\section{Direct Simulation Monte Carlo (DSMC) method}
\vspace{-0.2 cm}
We studied transport of vapor molecules leaving the evaporating liquid-vapor interface. Considering the vapor molecules behave as rigid rotators, vapor motion was obtained by numerical solution of BTE through the Direct Simulation Monte Carlo (DSMC) method. DSMC method is perfect for accurately and stably capturing the propagation of traveling discontinuities in the distribution function of BTE \cite{You2001,Peraud2014}. Furthermore, DSMC method is more computationally efficient compared to the other numerical methods based on discretization, both in terms of CPU time and storage. The collisions between water vapor molecules were handled using the variable soft sphere (VSS) collision model \cite{Lu} with a viscosity index $\omega$ = 1.047, scattering parameter $\alpha$ = 1.376, reference temperature $T_{ref}$ = 350 K, and reference molecular diameter $d_{ref}$)= 5.507 $\r{A}$. In each simulation, particles were weighted to represent numerous vapor molecules to reduce computational effort. The latest version of SPARTA (Stochastic PArallel Rarefied-gas Time-accurate Analyzer), an open-source DSMC program developed at Sandia National Laboratories was employed to solve BTE. In each simulation, the conditions were set identical to the experimental studies. We considered that the vapor molecules enter to the simulation box at temperature of $T_i^l$ reported in the experiments. The velocity of these vapor molecules were determined through centerline mass fluxes, $\dot{m}_{cl}$. The centerline mass fluxes were determined through the reported vapor and liquid heat fluxes at the centerline along with energy conservation law. Note that this mass flux is  local mass flux and is different than the average reported mass flux across the liquid-vapor interface. The vapor pressure in the domain of interest were set identical to the experimental conditions through adjusting the areal density of molecules in each simulation grid. The boundary condition at the top of the simulation box was set to the measured temperature $T_B$ at 3 mm above the interface. For few experimental studies \cite{DuanWardBadamDurst,DuanWard3}, temperature measurements in the vapor phase were not conducted up to 3 mm above the interface. To keep the coordinate of this boundary consistent between various studies, we extrapolated the reported vapor temperatures to 3 mm to find the value of $T_B$ in these few studies. Note that in these studies the vapor temperature gradient is constant and linear extrapolation is used.  Satisfying all the boundary conditions and initial thermodynamic properties in the domain of interest, we let the simulation run for $\tau$ seconds. $\tau$ is the time that a vapor molecule needs to travel from interface to the distance of z above the interface. The dimension of measurement probes (i.e. bead diameter) was 25 or 50 $\mu$m in all these studies, accuracy in spatial temperature measurements was $\pm$25$\mu$m (diameter of thermocouple), accuracy in spatial coordinate of liquid-vapor interface was $\pm$10 $\mu$m. Thus, we took $z$ equal to 85 $\mu$m as the upper boundary of all the experiments to ensures that the measurement probe was \textit{completely} in the vapor phase with no contact with the liquid phase during the experiments. We should add the determined vapor temperature through BTE at 85 $\mu$m was compared with measured temperature at 85 $\mu$m for all the studies. 

\begin{figure}[!ht]
 \begin{center}
\includegraphics[width=0.49 \textwidth]{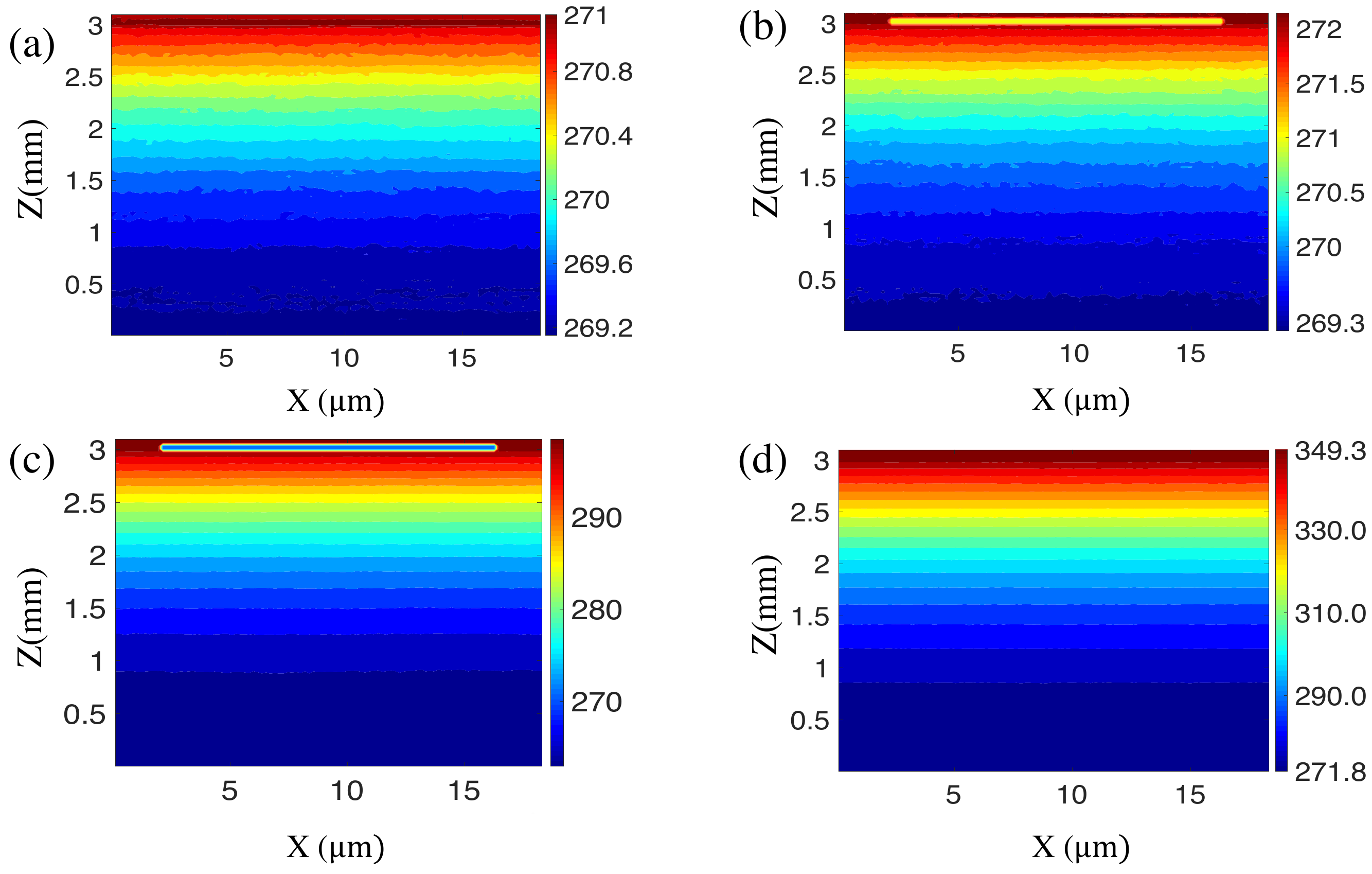}
\caption{Thermal map of vapor molecules obtained through DSMC simulations. These thermal maps are taken at time $\tau$ when the vapor molecules are at coordinate of 85 $\mu$m above the interface. The entering vapor molecules to the domain are at temperature of (a) 268.6 K, (b) 268.95 K, (c) 258.38 K and (d) 259.73 K.}
\label{Fig4}
\end{center}
\end{figure}

Figure \ref{Fig4} shows the calculated vapor temperature for four independent studies with the boundary conditions given in Fig. \ref{Fig2} after $\tau$ seconds. That is, in each simulation, vapor molecules enter to the domain with the initial temperature of $T_i^l$ with a given velocity determined through experimental centerline mass flux. As shown in these figures, temperature of vapor molecules changes as they transport through the domain. This temperature change is caused by the top imposed boundary temperature, $T_B$, in each experiment. The simulated vapor temperature along with those measured in a wide range of studies are tabulated in Table \ref{Table1}. Note that the difference in pressure is within the error bar of pressure measurements.We should add that these simulations at high vapor pressures are computationally expensive (e.g. each simulation takes 336 CPU hours on 100 processors). 

\begin{table}[]
\caption{Summary of all simulations at different vapor boundary conditions}
\begin{tabular}{cccccc}
\\
\\
\hline
\multicolumn{1}{c}{$\dot{m}_{cl}$} & \multicolumn{1}{c}{$P_{exp}$} & \multicolumn{1}{c}{$P_{sim}$} & \multicolumn{1}{c}{$T_{B}$} & \multicolumn{1}{c}{$T_i^v$(sim)} & \multicolumn{1}{c}{$T_i^v$(exp)} \\ 
\multicolumn{1}{c}{$\times$10$^{4}$ kg/(m$^2$s)} & \multicolumn{1}{c}{$\pm$ 13 Pa} & \multicolumn{1}{c}{Pa} & \multicolumn{1}{c}{K} & \multicolumn{1}{c}{K} & \multicolumn{1}{c}{K} \\ 
 \hline
 \multicolumn{6}{c}{Kazemi et al.\cite{Kazemi2017}}\\ 
\hline
3.97& 266 & 268 & 264.58 & 262.6 & 262.69 \\
3.88& 303 & 308 & 266.38 & 264.3 & 264.33 \\
3.08& 435 & 444 & 271.02 & 268.96 & 268.90 \\
2.35& 545 & 533 & 273.82 & 272.0 & 271.75 \\
0.65& 815 & 820 & 279.15 & 276.6 & 277.37 \\
\hline
\multicolumn{6}{c}{Jafari et al.\cite{Jafari2018}} \\ 
\hline
3.61& 374 & 373 & 270.15 & 267.15 & 267.15 \\
2.53& 436 & 430 & 272.15 & 269.2 & 269.15 \\
3.40& 526 & 520 & 273.95 &271.6 & 271.55 \\
2.24& 541 & 533 & 274.8 & 272.4 & 272.05 \\
2.18& 636 & 631 & 277.45 & 273.9 & 274.15 \\
0.72& 755 & 744 & 278.55 & 277.1& 276.55 \\
1.77& 913 & 911 & 281.4 & 279.0 & 279.25\\
\hline
\multicolumn{6}{c}{Fang and Ward \cite{FangWard} and Duan et al.\cite{DuanWard3}} \\ 
\hline
2.40& 194 & 198 & 286.38 & 268 & 266.40\\
0.87&196 & 193 & 300.07 & 263.1 & 263.67\\
0.56& 583 & 595 & 301.25 & 275.9 &275.25 \\
0.31& 591 & 602 & 294.73 & 275.5 &275.03\\
1.04& 625 & 630 & 302.97 & 275 & 275.33\\      
\hline 
\multicolumn{6}{c}{Badam et al.\cite{BadamKumarDurstDanov}}\\ 
\hline
7.20& 213 & 220 & 353.15 & 275.11& 275.40 \\
7.15& 288 & 288 & 353.15 & 275.5 & 277.80 \\
7.52& 388 & 401 & 353.15 & 280.5 & 280.92 \\
7.42& 569 & 565 & 353.15 & 283 & 284.00 \\
7.80& 744 & 780 & 353.15 & 287.8 & 286.83 \\
7.60& 855 & 894 & 353.15 & 287.5 & 288.05 \\
8.15& 946 & 972 & 353.15 & 288.3 & 289.10 \\
7.50& 1076 & 1090 & 353.15 & 291.1 & 291.00 \\
7.28& 215 & 210 & 343.15 & 273.15 & 274.78 \\
7.10& 290 & 295 & 343.15 & 274.5 & 276.40\\
6.76& 389 & 381 & 343.15 & 278 & 278.70 \\
6.5& 573 & 590 & 343.15 & 283 & 282.75 \\
6.91& 747 & 753 & 343.15 & 285.57 & 285.50 \\
6.96& 850 & 876 & 343.15 & 287.2 & 286.50 \\
 \hline
 \hline
\end{tabular}
\label{Table1}
\end{table}

\begin{figure}[!ht]
 \begin{center}
\includegraphics[width=0.49 \textwidth]{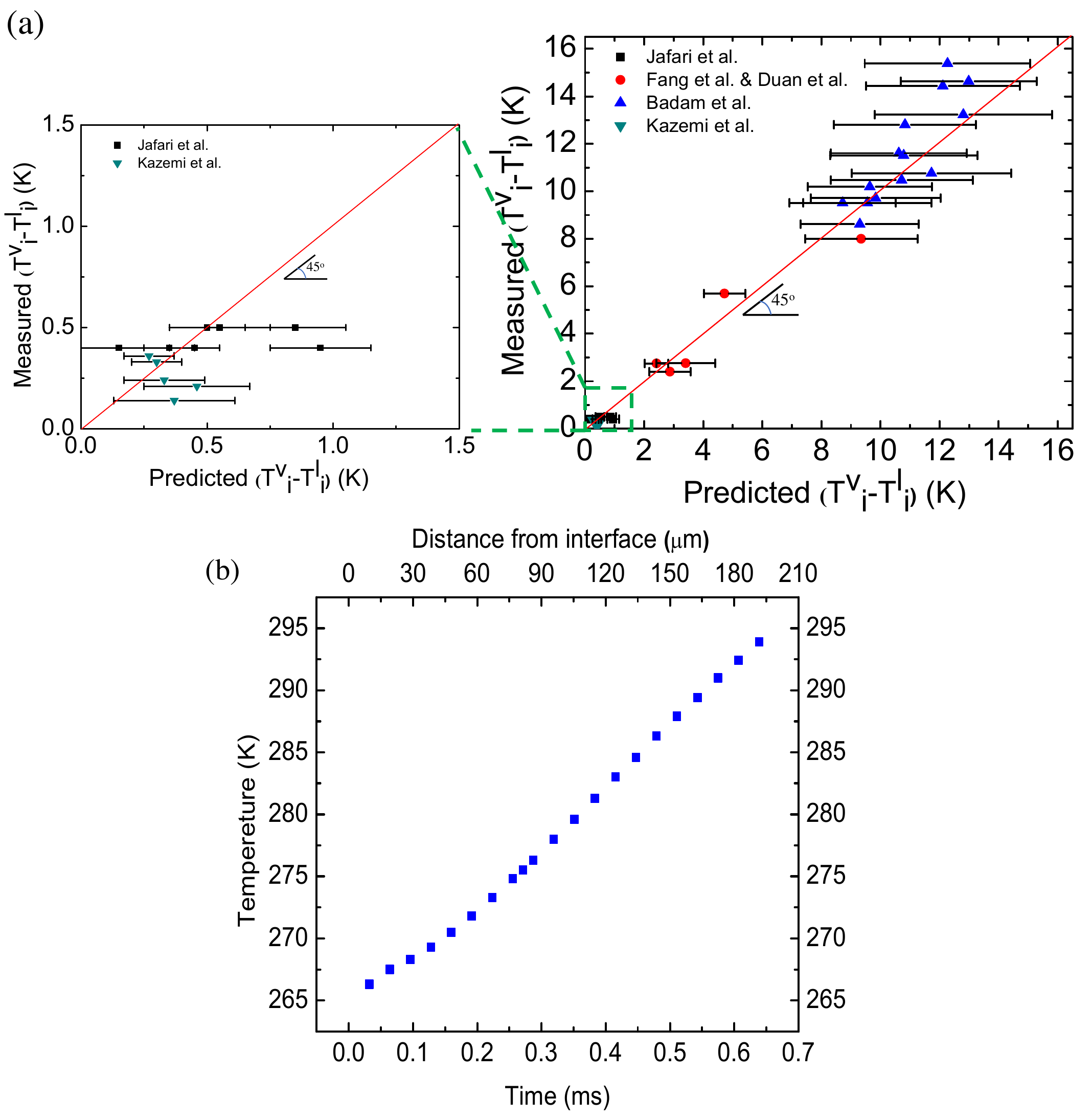}
\caption{(a) Computed temperature discontinuity by BTE is compared with the measured discontinuities in four independent groups. Note that thermal boundary conditions, $T_B$, varies in these studies. The error bars in these calculations are determined based on 17.5 $\mu$m error in vertical coordinate (i.e. half of 10 $\mu$m in position of liquid-vapor interface and 25 $\mu$m diameter of smallest thermocouple). (b) Temperature of vapor molecules leaving liquid-vapor interface changes as function of distance from the interface. Within 10 $\mu$m from the interface, temperature of vapor molecules has already changed by 2.91 K.}
\label{Fig5}
\end{center}
\end{figure}

The simulated values of temperature discontinuity at the interface are compared with the measured ones and are shown in Fig. \ref{Fig5}(a). As shown, there is a good agreement between the measured values and the simulated values. This is important as solution of BTE explains all the experimental findings from independent groups with good accuracy. Note that for closely similar vapor pressures and mass fluxes, the reported temperature discontinuities varied in two orders of magnitude (i.e. 0.14-28 K). This agreement indicates that the measured temperature discontinuities are strong function of imposed experimental boundary condition on the vapor phase, $T_B$. We think some deviations from 45 line is caused by inaccuracy in the measurements of $T_B$. As molecules leave the liquid-vapor interface, they are exposed to the hot temperature field and their temperature varies as they go further from the interface. The measured vapor temperature at any spatial coordinate above the interface only reflects the altered vapor temperature and is \textit{not} the interfacial vapor temperature (i.e. within few molecular length scale). Furthermore, this agreement supports the hypothesis that at these low evaporation rates, the assumption of $T_i^l$ $\approx$ $T_i^v$ is valid.To highlight this point, in Fig. \ref{Fig5}(b), temperature variation of vapor molecules is shown for the study by Badam et al. \cite{BadamKumarDurstDanov} in which vapor pressure was 288 Pa and mass flux was 11.9 $\times$10$^{-4}$ kg/(m$^2$s). In this study, $T_B$ (i.e. 3 mm above the interface) was set at 80 \textcelsius. The vapor molecules leave the liquid-vapor interface at temperature of 263.39 K. At the coordinate of 10 $\mu$m from the interface, temperature of vapor molecules has already changed to 266.3 K, which is 2.91 K higher than interfacial temperature. Note that 10 $\mu$m is far below the accuracy of measurements by thermocouples. Any temperature measurement by the thermocouples only reflects modified vapor temperature and not the interfacial vapor temperature. As molecules move further, their temperature could change by more than 30 K only within 500 $\mu$m. This finding also explain the reversed temperature discontinuity measured by Zhu et al. \cite{Zhu2009,Zhu2013a} in which vapor phase temperature was lower than liquid phase temperature. A possible approach to accurately measure interfacial vapor temperature is to minimize or suppress any heat flux by the vapor phase to the liquid-vapor interface. That is, experiments with lower vapor heat flux could provide better understanding of interfacial vapor temperature. The work by Kazemi et al. \cite{Kazemi2017} is the one with minimal vapor heat flux and indicates temperature discontinuity less than 0.14 $\pm$ 0.1 K at water evaporating interface. That is, at these low mass fluxes, the actual temperature discontinuity is $\le$ 0.1 K. We should add that this anomalous measured temperature discontinuity could occur for condensing vapor molecules on a liquid surface. 

In summary, through solution of BTE, we elucidated the source of controversial measured temperature discontinuities at an evaporating water interface. Although for closely similar conditions, the temperature discontinuity between 0.14 $\pm$ 0.1 to 28 $\pm$ 0.06 K are reported, these temperature measurements are strongly influenced by the imposed boundary condition on the vapor side and do not reflect the actual interfacial temperatures. As the molecules leaves the liquid-vapor interface, their temperature is changed in the vapor phase (e.g. 2.91 K within 10 $\mu$m). The conducted experiments with probe dimension of $\geq$25 $\mu$m could not provide an insight on the interfacial vapor temperature. A feasible way around this problem is to surpass heat flux on the vapor side to be able to extrapolate the measured vapor temperature to the interface. We believe that for these low mass fluxes, the temperature discontinuity across the liquid-vapor interface is so small. This understanding addresses this long-standing problem and provide a platform to further development of sound theories of evaporation. 

\begin{acknowledgments}
We wish to acknowledge the funding support by Air Force Office of Scientific Research (Grant AFOSR FA9550-16-1-0248) with Dr. Ali Sayir as program manager, National Science Foundation (Grant NSF- 1804204), Dr. Rodolfo Ostilla Monico for helpful discussions and Steve Plimpton and Moore Stan of Sandia National Laboratories for useful suggestions in SPARTA.
\end{acknowledgments}

\bibliographystyle{apsrev4-2}

%

\end{document}